\documentclass[showpacs,preprintnumbers,amsmath,amssymb,floatfix]{revtex4}

\usepackage{graphicx}
\usepackage{dcolumn}
\usepackage{bm}
\usepackage{epsfig}
\usepackage{feynmp}

\def\Bf#1{\mbox{\boldmath $#1$}}
\newcommand{\scs}{\scriptstyle}

\newcommand{\setval}{\fmfset{wiggly_len}{1.5mm}\fmfset{arrow_len}{1.5mm}\fmfset{arrow_ang}{13}\fmfset{dash_len}{1.5mm}\fmfpen{0.125mm}\fmfset{dot_size}{1thick}}
\begin{document}

\title{Large-$D$ Expansion from Variational Perturbation Theory}

\author{Sebastian F.\ Brandt}
\email{sbrandt@physics.wustl.edu}
\affiliation{%
Department of Physics, Campus Box 1105, Washington University in St.~Louis, MO 63130-4899, USA}%
\author{Axel Pelster}
\email{pelster@uni-essen.de}
\affiliation{%
Universit{\"a}t Duisburg-Essen, Campus Essen, Fachbereich Physik, Universit{\"a}tsstra{\ss}e 5, 45117
Essen, Germany}%

\date{\today}

\begin{abstract}
We derive recursively the perturbation series for the ground-state energy of the
$D$-dimensional anharmonic oscillator and resum it using variational perturbation theory (VPT).
From the exponentially fast converging approximants,
we extract the coefficients of the large-$D$ expansion to higher orders. 
The calculation effort is much
smaller than in the standard field-theoretic approach based on the 
Hubbard-Stratonovich transformation.
\end{abstract}
\pacs{02.30.Mv, 03.65.-w, 12.38.Cy}
\maketitle
\section{Introduction}
The properties of nontrivial
physical systems can only be calculated
via efficient approximation schemes.
Most easily accessible are
perturbation expansions, but they are usually divergent and need resummation.
To this end, a variational approach was developed by
Feynman and Kleinert \cite{Feynman2}, which has been systematically
extended to an efficient nonperturbative approximation scheme 
called {\it variational perturbation theory} (VPT)
\cite{Kleinertsys,PathInt3,VerenaBuch,Festschrift}. It allows the conversion
of divergent weak-coupling into convergent
strong-coupling expansions and has been applied successfully in various fields, such as
quantum mechanics, quantum
statistics, condensed matter physics, and
critical phenomena. In fact, the most accurate critical exponents
come from this theory \cite{seven},
as has been verified by recent satellite experiments \cite{LIPA}.\\

The convergence of VPT has been analyzed up to very high orders for the ground-state energy
of the one-dimensional anharmonic oscillator
\begin{eqnarray}
\label{1D}
V(x) = \frac{1}{2}\,\omega^2 x^2 + g x^4
\end{eqnarray}
and was found to be exponentially fast \cite{JankeC1,JankeC2}.
This surprising result has been confirmed later by studying other physical systems and was
proven to hold on general grounds \cite{PathInt3,VerenaBuch}. Furthermore, the
exponential convergence seems to be uniform with respect to other system parameters.
In this manner, the variational resummation of perturbation series yields approximations
which are generically reasonable for all temperatures \cite{Meyer,Werner,Weissbach},
space and time coordinates \cite{Michael1,Putz,Schanz,Dreger},
magnetic field strengths \cite{Michael2}, 
coupling constants \cite{C1,C2,Brandt}, etc. \\

In this paper, we show that the exponential convergence of VPT is uniform
with respect to the space dimension $D$. To this end, we consider the $D$-dimensional
generalization of the anharmonic oscillator (\ref{1D}), i.e.
\begin{eqnarray}
\label{V}
V({\bf x}) = \frac{1}{2}\, \omega^2 {\bf x}^2 + g \left( {\bf x}^2 \right)^2
\end{eqnarray}
with ${\bf x}=(x_1,\ldots,x_D)$, and determine its ground-state energy as a function of the
coupling constant $g$. In Section II, we derive the corresponding weak-coupling series by
evaluating connected vacuum diagrams. In Section III, we show how this perturbation series
can be obtained more efficiently by
means of Bender-Wu-like recursion relations \cite{Bender/Wu}. In Section IV, we resum the
weak-coupling series by applying VPT and examine the resulting convergence, which is again
exponentially fast and improves uniformly
with increasing dimension $D$. In Section V, we show that the
latter observation is not surprising, since the ground-state energy of the anharmonic
oscillator (\ref{V}) can be determined with the help of
a systematic large-$D$ expansion. In Section VI, we apply VPT to extract the large-$D$
expansion to higher orders, which have so far been inaccessible to other methods.
\section{Perturbation Theory}
\begin{fmffile}{graph}
\setlength{\unitlength}{1mm}
The perturbation series
for the ground-state energy of the anharmonic oscillator (\ref{V})
can be calculated from connected vacuum diagrams. Up to the second
order in the coupling constant $g$, the ground-state energy is given
by the Feynman diagrams
\begin{eqnarray}
E =  \frac{\omega}{2} - \lim_{\beta \to \infty} \frac{1}{\beta} \hspace*{1mm} \bigg\{
\hspace*{1mm}
\frac{1}{8}
\hspace*{1mm}
\parbox{11mm}{\centerline{
\begin{fmfgraph}(8,4)
\setval
\fmfleft{i1}
\fmfright{o1}
\fmf{plain,left=1}{i1,v1,i1}
\fmf{plain,left=1}{o1,v1,o1}
\fmfdot{v1}
\end{fmfgraph}}}
+ \frac{1}{48}
\hspace*{1mm}
\parbox{9mm}{\centerline{
\begin{fmfgraph}(6,4)
\setval
\fmfforce{0w,0.5h}{v1}
\fmfforce{1w,0.5h}{v2}
\fmf{plain,left=1}{v1,v2,v1}
\fmf{plain,left=0.4}{v1,v2,v1}
\fmfdot{v1,v2}
\end{fmfgraph}}}
\hspace*{1mm}
+\frac{1}{16}
\hspace*{1mm}
\parbox{15mm}{\centerline{
\begin{fmfgraph}(12,4)
\setval
\fmfleft{i1}
\fmfright{o1}
\fmf{plain,left=1}{i1,v1,i1}
\fmf{plain,left=1}{v1,v2,v1}
\fmf{plain,left=1}{o1,v2,o1}
\fmfdot{v1,v2}
\end{fmfgraph}}} + \ldots \bigg\} \,,
\label{dia3l}
\end{eqnarray}
with the propagator
\begin{eqnarray}
\parbox{7mm}{\centerline{
  \begin{fmfgraph*}(14,10)
  \setval
  \fmfforce{0.w,0.5h}{v1}
  \fmfforce{0.w,1h}{v11}
  \fmfforce{0.w,0h}{v12}
  \fmfforce{1w,0.5h}{v2}
  \fmfforce{1w,1h}{v21}
  \fmfforce{1w,0h}{v22}
  \fmf{plain}{v1,v2}
  \fmfv{decor.size=0, label=${\scs a}$, l.dist=1mm, l.angle=-180}{v1}
  \fmfv{decor.size=0, label=${\scs b}$, l.dist=1mm, l.angle=0}{v2}
  \fmfv{decor.size=0, label=${\scs i}$, l.dist=2.9mm, l.angle=50}{v12}
  \fmfv{decor.size=0, label=${\scs j}$, l.dist=2.5mm, l.angle=130}{v22}
  \fmfdot{v1}
  \fmfdot{v2}
  \end{fmfgraph*} } }
\hspace*{8mm}
\equiv \hspace*{2mm} \frac{\delta_{ij}}{2\omega}e^{-\omega|\tau_a - \tau_b|}
\end{eqnarray}
and the vertices
\begin{eqnarray}
 \parbox{7mm}{\centerline{
  \begin{fmfgraph*}(6,6)
  \setval
  \fmfforce{0w,0.5h}{v1}
  \fmfforce{0.5w,1h}{v2}
  \fmfforce{1w,0.5h}{v3}
  \fmfforce{0.5w,0h}{v4}
  \fmfforce{1/2w,1/2h}{v5}
  \fmf{plain}{v1,v3}
  \fmf{plain}{v2,v4}
  \fmfv{decor.size=0, label=${\scs i}$, l.dist=1mm, l.angle=-180}{v1}
  \fmfv{decor.size=0, label=${\scs j}$, l.dist=1mm, l.angle=90}{v2}
  \fmfv{decor.size=0, label=${\scs k}$, l.dist=1mm, l.angle=0}{v3}
  \fmfv{decor.size=0, label=${\scs l}$, l.dist=1mm, l.angle=-90}{v4}
  \fmfv{decor.size=0, label=${\scs a}$, l.dist=1mm, l.angle= 45}{v5}
  \fmfdot{v5}
  \end{fmfgraph*} } }
\hspace*{5mm}
\equiv \hspace*{2mm} - \frac{g}{3} \hspace*{1mm} \bigg\{ \delta_{ij}\delta_{kl}
+ \delta_{ik}\delta_{jl} + \delta_{il}\delta_{jk} \bigg\}
\int_0^{\beta} d\tau_a \,.
\end{eqnarray}
The connected vacuum diagrams (\ref{dia3l}) are derived most easily by an efficient
graphical recursion method \cite{Rek}.
Evaluating these Feynman diagrams produces the following analytic
expression for the ground-state energy:
\begin{eqnarray}
E = \frac{D\omega}{2} + \frac{D(D+2)g}{4\omega^2} -
\frac{D(2D^2+9D+10)g^2}{8\omega^{5}}  + \ldots
\label{PerExpigx3} \,.
\end{eqnarray}
Only low perturbation orders are accessible by this procedure. If we want
to study higher orders, we better use  Bender-Wu-like recursion relations
\cite{Bender/Wu}.
\section{Bender-Wu-Like Recursion Relations}
The potential (\ref{V}) of the $D$-dimensional anharmonic oscillator
is rotationally symmetric. Hence, the ground-state wave function
depends only on the distance $x = |{\bf x}|$. We solve the corresponding Schr{\"o}dinger
eigenvalue equation
\begin{eqnarray}
\label{S}
\left[ - \frac{1}{2}\left(\frac{\partial^2}{\partial x^2} + \frac{D-1}{x}\frac{\partial}{\partial x}\right)  + \frac{1}{2} \,\omega^2 x^2 + g x^4 \right]
\psi(x) = E \,\psi(x)
\end{eqnarray}
as follows. We write the wave function $\psi(x)$ as
\begin{eqnarray}
\psi(x) =  \left(\frac{\omega}{\pi}\right)^{1/4} \exp
\left[ - \frac{1}{2} \,\hat{x}^2 + \phi(\hat{x}) \right]\,,
\label{CAns}
\end{eqnarray}
with the abbreviation $\hat{x} = x\sqrt{\omega}$,
and expand the exponent in powers of the dimensionless coupling constant
$\hat{g} = g / \omega^3$ by using
\begin{eqnarray}
\phi(\hat{x}) = \sum_{k = 1}^{\infty} \phi_k(\hat{x})\hat{g}^k  \,.
\label{PhiAns}
\end{eqnarray}
The $\phi_k(\hat{x})$ are expanded in powers of the rescaled coordinate $\hat{x}$:
\begin{eqnarray}
\label{EX}
\phi_k(\hat{x}) = \sum_{m=1}^{k+1} c_{m}^{(k)} \hat{x}^{2m} \, .
\end{eqnarray}
For the ground-state energy we make the ansatz
\begin{eqnarray}
E = \omega \left(  \frac{D}{2} + \sum_{k = 1}^{\infty}  \epsilon_k \hat{g}^k \right) \,.
\label{AnsE0D}
\end{eqnarray}
Inserting (\ref{CAns})--(\ref{AnsE0D}) into (\ref{S}), we obtain to first order
\begin{eqnarray}
c_1^{(1)} = - \frac{D+2}{4} \, , \hspace*{0.3cm}
c_2^{(1)} = - \frac{1}{4} \, , \hspace*{0.3cm}
\epsilon_1 = \frac{D(D + 2)}{4} \, .
\label{eps2Ddim}
\end{eqnarray}
For $k \geq 2$, we find the following recursion relation for the expansion
coefficients in (\ref{EX})
\begin{eqnarray}
c_m^{(k)} =  \frac{(m+1)(D+2m)}{2m} \,c_{m + 1}^{(k)}
+ \sum_{l = 1}^{k-1}\sum_{n = 1}^{m} \frac{n(m+1-n)}{m} \, c_n^{(l)}\, c_{m +1 - n}^{(k - l)}\,,
\label{REK}
\end{eqnarray}
with $c_m^{(k)} \equiv 0$ for $m > k + 1$. The expansion coefficients of the
ground-state energy follow from
\begin{eqnarray}
\epsilon_k = - D \, c_1^{(k)} \, .
\label{ekDnodim}
\end{eqnarray}
Table \ref{EpsTableD} shows the resulting
coefficients $\epsilon_k$ up to the fifth order. For $D=1$, they reduce
to the well-known one-dimensional results \cite{Bender/Wu}.
\begin{table}[t]
\begin{center}
\begin{tabular}{|r|c|}
\hline
\rule[-1mm]{0mm}{4mm}\hspace*{1mm}$k$\hspace*{1mm}& $\epsilon_k$ \\[1mm] \hline
\rule[-1mm]{0mm}{5mm}\hspace*{1mm}$1$\hspace*{1mm} & ${\displaystyle D(D + 2) / 4 }$
\\[1.5mm] \hline
\rule[-1mm]{0mm}{5mm}\hspace*{1mm}$2$\hspace*{1mm}& ${\displaystyle - D(2D^2
+9D+10)/ 8 }$ \\[1.5mm] \hline
\rule[-1mm]{0mm}{5mm}\hspace*{1mm}$3$\hspace*{1mm}& $ {\displaystyle D(8 D^3
+59D^2 +  146D + 120   ) / 16 }$ \\ [1.5mm]\hline
\rule[-1mm]{0mm}{5mm}\hspace*{1mm}$4$\hspace*{1mm} & $ \hspace*{1mm}{\displaystyle -D( 168D^4
+   1773 D^3 + 7144D^2 +  12960D + 8840  ) / 128  }$ \hspace*{1mm}\\[1.5mm] \hline
\rule[-1mm]{0mm}{5mm}\hspace*{1mm}$5$\hspace*{1mm} & $ \hspace*{1mm}{\displaystyle D( 1024D^5
+ 14325 D^4 + 82222 D^3
+ 241464 D^2 + 360736 D + 216960 )/256 }$ \hspace*{1mm}
\\[1.5mm] \hline
\end{tabular}
\caption{Expansion coefficients for the ground-state energy of the
  anharmonic oscillator (\ref{V}) up to the fifth order.}
\label{EpsTableD}
\end{center}
\end{table}
\section{Variational Resummation}
Now we consider the strong-coupling limit of the
perturbation series (\ref{AnsE0D}). Rescaling the
coordinate according to $x \to xg^{-1/6}$,
the Schr{\"o}dinger equation (\ref{S}) becomes
\begin{eqnarray}
\hspace*{-3mm} \left[ - \frac{1}{2} \left(\frac{\partial^2}{\partial x^2} + \frac{D-1}{x}\frac{\partial}{\partial x}\right) +  \frac{1}{2} \,g^{-2/3} \omega^2 x^2 + x^4
\right] \psi(x) = g^{-1/3}\,E \, \psi(x) \, .
\label{SchrEqReSca}
\end{eqnarray}
Expanding the wave function and the energy in powers of the coupling constant yields
\begin{eqnarray}
\label{EX1}
\hspace*{-0.4cm} \psi(x) &=& \psi_0(\hat{x}) + \psi_1(\hat{x}) \,\hat{g}^{-2/3}
+ \psi_2(\hat{x}) \, \hat{g}^{-4/3}+ \ldots \, , \\
\hspace*{-0.4cm} E &=&  \omega  \,\hat{g}^{1/3}
\left(b_0 +  b_1 \,\hat{g}^{-2/3}+ b_2 \,\hat{g}^{-4/5}+ \ldots \right) \, .
\label{EX2}
\end{eqnarray}
By considering (\ref{SchrEqReSca}) in the limit $g\to \infty$,
we find that the leading strong-coupling coefficient $b_0$ equals
the ground-state energy associated with the  Hamilton operator
\begin{eqnarray}
\label{HAM}
H = - \frac{1}{2} {\Bf \Delta} + \left( {\bf x}^2 \right)^2 \, .
\end{eqnarray}
Precise numerical values for this ground-state energy for different dimensions $D$ are listed
in Table \ref{TABHAM}.\\

The weak-coupling series (\ref{AnsE0D}) is of the form
\begin{eqnarray}
E^{(N)}(g , \omega) = \omega \left[ \frac{D}{2}+ \sum_{k = 1}^N
 \epsilon_{k} \left( \frac{g}{\omega^3} \right)^k \right]  \, .
\label{Eexp}
\end{eqnarray}
The alternating signs and fast growing coefficients in
Table \ref{EpsTableD} suggest that (\ref{Eexp}) is a divergent Borel series which is
resummable by VPT \cite{Kleinertsys,PathInt3,VerenaBuch,Festschrift}.
To this end, an artificial frequency parameter $\Omega$ is introduced in the perturbation
series most easily by Kleinert's square-root trick
\begin{eqnarray}
\omega \to \Omega \sqrt{1 + g  r}\,,
\label{KleTrick1}
\end{eqnarray}
with
\begin{eqnarray}
r = \frac{\omega^2 - \Omega^2}{g\Omega^2}\,.
\label{KleTrick2}
\end{eqnarray}
One replaces the frequency $\omega$ in the weak-coupling series
(\ref{Eexp}) according to (\ref{KleTrick1})
and re-expands the resulting expression in powers of $g$ up to the order
$g^{N}$. Afterwards, the parameter $r$ is replaced according to (\ref{KleTrick2}).
This procedure has the effect that the power series of order $N$ for the
ground-state energy becomes dependent on
the variational parameter $\Omega$:
\begin{eqnarray}\hspace*{-0.9cm}
E^{(N)}(g, \omega, \Omega) &=& \sum_{k = 0}^{N} \epsilon_k g^k \Omega^{1 - 3 k}
\sum_{l = 0}^{N - k} {(1 - 3 k)/2 \choose  l} \left( \frac{\omega^2}{\Omega^2} - 1 \right)^l \,.
\label{fNgK}
\end{eqnarray}
The influence of $\Omega$ is then optimized according to the principle of minimal
sensitivity \cite{Stevenson}: the ground-state energy to $N$th order is approximated by
\begin{eqnarray}
E^{(N)} = E^{(N)}(g, \omega, \Omega^{(N)}) \,,
\end{eqnarray}
where $\Omega^{(N)}$ denotes that value of the variational parameter for
which $E^{(N)}(g, \omega, \Omega)$ has an extremum or a turning point.\\

\begin{table}[t]
\begin{center}
\begin{tabular}{|l|l|}
\hline
\rule[-1mm]{0mm}{4.5mm}\hspace*{1mm}$b_0(D = 2)$\hspace*{1mm} &
\hspace*{1mm}1.4771497535779972(31) \hspace*{1mm} \\[1mm] \hline
\rule[-1mm]{0mm}{4.5mm}\hspace*{1mm}$b_0(D = 3)$\hspace*{1mm}&
\hspace*{1mm}2.3936440164822970(37) \hspace*{1mm} \\[1mm] \hline
\rule[-1mm]{0mm}{4.5mm}\hspace*{1mm}$b_0(D = 10)$\hspace*{1mm}&
\hspace*{1mm}10.758265165443755(69) \hspace*{1mm}\\[1mm] \hline
\end{tabular}
\caption{Numerical results for the leading strong-coupling
coefficient $b_0$ for the ground-state energy of (\ref{HAM}).}
\label{TABHAM}
\end{center}
\end{table}
\begin{table}[t]
\begin{center}
\begin{tabular}{|l|l|}
\hline
\rule[-1mm]{0mm}{4.5mm}\hspace*{1mm}$b_0(D = 2)$\hspace*{1mm}  &
\hspace*{1mm}1.477149753577994356(33)\hspace*{1mm}  \\[1mm] \hline
\rule[-1mm]{0mm}{4.5mm}\hspace*{1mm}$b_0(D = 3)$\hspace*{1mm}  &
\hspace*{1mm}2.3936440164823030895(77)\hspace*{1mm}  \\[1mm] \hline
\rule[-1mm]{0mm}{4.5mm}\hspace*{1mm}$b_0(D = 10)$\hspace*{1mm} &
\hspace*{1mm}10.758265165443797408091(18)\hspace*{1mm}  \\[1mm] \hline
\end{tabular}
\caption{High-precision VPT results from the 80th order
for the leading strong-coupling coefficient $b_0$
in the ground-state energy (\ref{V}).}
\label{ResultsD}
\end{center}
\end{table}
As an example, consider the weak-coupling series (\ref{Eexp}) to first order:
\begin{eqnarray}
E^{(1)} = \frac{D \omega}{2} + \frac{D(D + 2)}{4 \omega^2}\,g\,.
\label{E1OmD}
\end{eqnarray}
Inserting (\ref{KleTrick1}), re-expanding in $g$ to first order, and
taking into account (\ref{KleTrick2}), we find
\begin{eqnarray}
\label{E1}
E^{(1)}(g,\omega,\Omega) = \frac{D \Omega}{4} +  \frac{D \omega^2}{4 \Omega} +
\frac{D(D+2)}{4 \Omega^2}\,g\,.
\end{eqnarray}
Extremizing this and going to large coupling constants, we obtain the strong-coupling behavior
of the variational parameter
\begin{eqnarray}
\hspace*{-4mm} \Omega^{(1)} = \omega\, \hat{g}^{1/3}\left(\Omega_0^{(1)} + \Omega_1^{(1)}\hat{g}^{-2/3}
+\Omega_2^{(1)}\hat{g}^{-4/3} + \ldots \right) \, ,
\label{Om1Da}
\end{eqnarray}
with
\begin{eqnarray}
\Omega_0^{(1)} = \sqrt[3]{2(D+2)} \,, \hspace*{1cm}
\Omega_1^{(1)} = \frac{\omega^2}{3 \sqrt[3]{2(D+2)} }\,, \hspace*{1cm}
\Omega_2^{(1)} = \frac{\omega^4}{108(D+2)}
\, ,  \hspace*{0.2cm}\ldots \, .
\label{Om1Db}
\end{eqnarray}
Inserting the result (\ref{Om1Da}), (\ref{Om1Db}) into
(\ref{E1}), we obtain the strong-coupling
series (\ref{EX2}) with the first-order coefficients
\begin{eqnarray}
\hspace*{-3mm} b_0^{(1)} = \frac{3D}{8} \sqrt[3]{2(D+2)} \,, \hspace*{1cm}
b_1^{(1)} = \frac{D \omega^2}{4\sqrt[3]{2(D+2)}} \,, \hspace*{1cm}
b_2^{(1)} = - \frac{D \omega^4}{48 (D+2)}
\, ,\hspace*{0.2cm} \ldots \,.
\label{b1D}
\end{eqnarray}
For $D=1$, this reduces to earlier VPT results in Refs.~\cite{JankeC1,JankeC2}. \\

In order to determine the strong-coupling coefficient $b_0$ in (\ref{EX2}) to higher
orders, we proceed as follows. We observe that the strong-coupling behavior
of the variational parameter is for any order $N$ of the form
\begin{eqnarray}
\hspace*{-3mm} \Omega^{(N)} = \omega \hat{g}^{1/3}\left(\Omega_0^{(N)} + \Omega_1^{(N)}\hat{g}^{-2/3}
+\Omega_2^{(N)}\hat{g}^{-4/3} + \ldots \right)
\label{Om1DaB}
\end{eqnarray}
which corresponds to (\ref{Om1Da}).
Inserting (\ref{Om1DaB}) into (\ref{fNgK}), the leading strong-coupling coefficient turns
out to be given by
\begin{eqnarray}
b_0^{(N)} \left( \Omega_0^{(N)} \right) &=& \sum_{k = 0}^N\sum_{l = 0}^{N - k} { (1
  -3 k)/2 \choose l} (-1)^l \epsilon_k(\Omega_0^{N})^{1 - 3 k} \,,
\end{eqnarray}
where the inner sum can be further simplified by using \cite{Gradshteyn}
\begin{eqnarray}
\sum_{l = 0}^m(-1)^l{\alpha \choose l} = (-1)^m{ \alpha -1 \choose m} \,.
\end{eqnarray}
Thus, the leading strong-coupling coefficient reduces to
\begin{eqnarray}
b_0^{(N)}(\Omega_0^{(N)}) &=& \sum_{k =0}^N
{(1 -3 k )/2 - 1 \choose N - k} (-1)^{N-k} \epsilon_k\left(\Omega_0^{(N)}\right)^{1 -3 k} \,.
\label{b0Nred}
\end{eqnarray}
Determining the optimized $\Omega_0^{(N)}$ for which $b_0^{(N)}(\Omega_0^{(N)})$ has an
extremum or a turning point then leads to the approximation $b_0^{(N)}(\Omega_0^{(N)})$
of the leading strong-coupling coefficient $b_0$.\\

\begin{figure}[t]
\centerline{\includegraphics[scale=0.65]{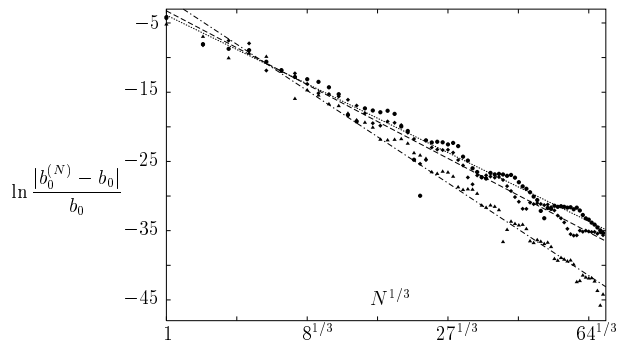}}
\caption{Logarithm of the relative deviation of the
VPT result for the leading strong-coupling coefficient $b_0(D)$ for $D=2$
(circles), $D=3$ (diamonds), and $D=10$ (triangles) plotted as a function of
the cubic root of the perturbation order up to 70th
order. The dashed lines represent least squares fits of the data to straight lines.}
\label{ComD}
\end{figure}

By carrying the expansion to high orders,
VPT yields approximations whose relative deviation from the exact value vanishes
exponentially \cite{PathInt3,VerenaBuch}.
In our case, we have for large $N$
\begin{eqnarray}
\frac{|b_0^{(N)} (D)- b_0(D)|}{b_0(D)} \approx \exp\left[A(D) - B(D) N^{1/3} \right]\,,
\label{KonVer}
\end{eqnarray}
where the exponent $1/3$ is determined by the structure of the
strong-coupling series (\ref{EX2}).
Due to the exponential convergence of VPT, it turns out that the
accuracy of our numerical results for the leading
strong-coupling coefficients $b_0$ in Table \ref{TABHAM} is not sufficient for a useful
examination of the
convergence behavior of VPT in high orders. Therefore, we use our results from
the $80$th VPT order as a more precise approximation for $b_0$. Table \ref{ResultsD}
summarizes our high-precision VPT results, where in each case the uncertainty of $b_0$ has been
estimated by examining the deviation from the result of the previous
order. For $D=2$ and $D=10$, the VPT results lie within the error
margins of the numerical results. However, this is not the case for $D
= 3$, where the VPT result lies just outside of the corresponding
numerical error margins. We attribute this discrepancy to an overly
optimistic error estimate for the numerical result. The precision of the
results shown in
Table \ref{ResultsD} improves with increasing dimension, which already
indicates that the calculation converges faster in higher dimensions. Figure
\ref{ComD} shows the convergence of the  VPT results for the three different
cases. Fitting the data to straight lines yields for the parameters $A(D)$ and $B(D)$
in (\ref{KonVer}):
\begin{eqnarray}
\hspace*{-6mm} A(D=2) &=& 5.98(72) \, , \hspace*{0.2cm} B(D=2)= 9.89(23) \, , \\
\hspace*{-6mm} A(D=3) &=& 7.43(48) \, , \hspace*{0.2cm} B(D=3)= 10.67(15)\, , \\
\hspace*{-6mm} A(D=10) &=& 11.89(63)\, , \hspace*{0.2cm} B(D=10)=13.33(20) \, .
\end{eqnarray}
Thus, we find that the convergence of the VPT result indeed improves uniformly with
increasing dimension $D$. This tendency does not come as a surprise,
since for $D \to \infty$ the ground-state energy of an oscillator with quartic anharmonicity
can be determined exactly as we will see in the subsequent section.

\section{Large-$D$ Expansion}
Now, we elaborate the systematic large-$D$ expansion for the ground-state energy of the
anharmonic oscillator (\ref{V}) based on standard
field-theoretic methods (see, for instance,
Refs.~\cite{PathInt3,Zinn,Brezin}).
\subsection{Effective Potential}
We start with the path integral representation of the
quantum-statistical partition function at finite temperature  $T$
\begin{eqnarray}
\label{PF}
Z = \oint {\cal D} {\bf x} \, e^{- {\cal A}[{\bf x}]} \, ,
\end{eqnarray}
where the Euclidean action reads
\begin{eqnarray}
\label{AC1} \hspace*{-2mm}
{\cal A} [{\bf x}] = \int_0^{\beta} d \tau \left\{ \frac{1}{2} \,\dot{\bf x}^2 ( \tau ) +
\frac{1}{2} \,\omega^2 {\bf x}^2 (\tau)+ g \left[ {\bf x}^2 (\tau)\right]^2 \right\} \, .
\end{eqnarray}
The paths are periodic in the imaginary time $\tau$ with period $\beta \equiv 1/T$.
Applying a Hubbard-Stratonovich transformation,
\begin{eqnarray}
\oint {\cal D}\sigma \exp \left\{ -  \int_0^{\beta} d\tau \, \left[\frac{1}{g}\,
\sigma^2(\tau) + 2i {\bf x}^2(\tau) \sigma (\tau) \right] \right\}
 = \exp \left\{ - g \int_0^{\beta} d\tau \, \left[{\bf x}^2(\tau) \right]^2 \right\} \, ,
\label{HS}
\end{eqnarray}
the ${\bf x}(\tau)$-path integral (\ref{PF}), (\ref{AC1}) becomes harmonic and leads to
\begin{eqnarray}
Z = \oint {\cal D} \sigma \, e^{- D {\cal A}[\sigma] } \, ,
\label{SIG}
\end{eqnarray}
where we have introduced the Euclidean action
\begin{eqnarray}
{\cal A}[\sigma] = \frac{1}{D g} \int_0^{\beta} d \tau \,\sigma^2(\tau) + \frac{1}{2}\,
\mbox{Tr}\,  \ln\left[-\frac{d^2}{d\tau^2} + \omega^2 + 4i  \sigma (\tau) \right] \, .
\label{AC2}
\end{eqnarray}
The remaining path integral (\ref{SIG}), (\ref{AC2})  over $\sigma (\tau )$
is then performed in the limit of large $D$ by regarding the modified
coupling constant $\tilde{g}=D g$ as being independent of $D$.
and by applying the background method \cite{PathInt3,DeWitt,Jackiw}.
Thus, we take into account order by order the effect of the fluctuations
$\delta \sigma ( \tau)\equiv \sigma(\tau)- \sigma_0$ of the paths $\sigma ( \tau )$ from the background $\sigma_0$.
This determines the effective potential
\begin{eqnarray}
V_{\rm eff} ( \sigma_0 ) = - \frac{1}{\beta}\,\ln Z
\end{eqnarray}
in the form of the loop expansion
\begin{eqnarray}
V_{\rm eff}(\sigma_0) = \sum_{l=0}^{\infty} V_{\rm eff}^{(l)}(\sigma_0) \, ,
\end{eqnarray}
where the term of loop order $l$ turns out to be of order $D^{1-l}$.
\subsection{Loop Orders $l=0$ and $l=1$}
The leading term is the tree-level and  follows from evaluating (\ref{AC2})
for the background $\sigma_0$
\begin{eqnarray}
\label{L0}
V_{\rm eff}^{(0)}(\sigma_0) = D \left[ \frac{\sigma_0^2}{\tilde{g}} +
\frac{1}{\beta} \ln \left( \sinh  \frac{\beta \Omega}{2} \right) \right]
\end{eqnarray}
with the auxiliary frequency
\begin{eqnarray}
\label{FR}
\Omega  = \sqrt{\omega^2 + 4 i \sigma_0 } \, .
\end{eqnarray}
The  one-loop correction is given by
\begin{eqnarray}
\label{TR1}
V_{\rm eff}^{(1)}(\sigma_0) = \frac{1}{2 \beta}\,\mbox{Tr}\, \ln G^{-1} (\tau_1,\tau_2) \, ,
\end{eqnarray}
where the operator
\begin{eqnarray}
G^{-1} (\tau_1,\tau_2) = \left. \frac{\delta^2 {\cal A}[\sigma]}{\delta \sigma(\tau_1) \delta \sigma(\tau_2)} \right|_{\sigma(\tau)=\sigma_0}
\end{eqnarray}
turns out to be
\begin{eqnarray}
\label{GGG}
G^{-1} (\tau_1,\tau_2) =\frac{2}{\tilde{g}} \, \delta(\tau_1 - \tau_2 ) + 8\,G_\Omega^2 (\tau_1,\tau_2 )  \, .
\end{eqnarray}
Here, the correlation function $G_\Omega (\tau_1,\tau_2 )$ has the spectral representation
\begin{eqnarray}
\label{CF}
G_\Omega (\tau_1,\tau_2 ) =  \sum_{m=-\infty}^\infty
\frac{e^{-i \omega_m (\tau_1 - \tau_2 )}}{\beta (\omega_m^2 + \Omega^2)} \, ,
\end{eqnarray}
with the Matsubara frequencies $\omega_m=2 \pi m/ \beta$. Inserting (\ref{CF}) in (\ref{GGG}) yields
\begin{eqnarray}
\label{OM}
G^{-1} (\tau_1,\tau_2) = \frac{1}{\beta}\sum_{m=-\infty}^\infty G_m^{-1} \, e^{-i \omega_m (\tau_1 - \tau_2 )} \, ,
\end{eqnarray}
with the coefficients
\begin{eqnarray}
\label{MC}
G_m^{-1} = \frac{2}{\tilde{g}} +
\frac{8}{\beta} \sum_{m'=-\infty}^\infty \frac{1}{(\omega_{m'}^2+\Omega^2) (\omega_{m-m'}^2+\Omega^2)} \, .
\end{eqnarray}
For the quantum-mechanical ground state energy to be calculated
we only need the zero-temperature limit $\beta \to \infty$ of the above 
quantum statistical expressions. Thus, we obtain from (\ref{L0})
\begin{eqnarray}
\label{L0B}
V_{\rm eff}^{(0)}(\sigma_0) \to  D \left( \frac{\sigma_0^2}{\tilde{g}} + \frac{\Omega}{2} \right),
\end{eqnarray}
and the Matsubara sum (\ref{MC}) reduces to an integral,
\begin{eqnarray}
\label{CON}
\sum_{m=-\infty}^\infty f(\omega_m) \to \frac{\beta}{2 \pi}
\int_{-\infty}^\infty d \omega_m \,
f(\omega_m) \, ,
\end{eqnarray}
yielding
\begin{eqnarray}
\label{MCC}
G_m^{-1} = \frac{2}{\tilde{g}} + \frac{8}{\Omega (\omega_m^2 + 4 \Omega^2)} \, .
\end{eqnarray}
Correspondingly, Eq.~(\ref{TR1}) becomes
\begin{eqnarray}
\label{TR2}
V_{\rm eff}^{(1)}(\sigma_0) =
\frac{1}{2 \beta}\, \sum_{m=-\infty}^\infty \ln  G_m^{-1}\to
\frac{1}{4\pi}\,
\int_{-\infty}^\infty d \omega_m \,
\ln  G_m^{-1} \,
= \frac{\tilde{\Omega}}{2} - \Omega \, ,
\end{eqnarray}
where
\begin{eqnarray}
\label{AUX2}
\tilde{\Omega} = 2\,\sqrt{\Omega^2 + \frac{\tilde{g}}{\Omega}}
\end{eqnarray}
denotes another auxiliary frequency.
\subsection{Loop Order $l=2$}
The higher loop orders of the effective potential with $l \geq 2$ consist of all one-particle
irreducible vacuum diagrams with the propagator
\begin{eqnarray}
  \parbox{7mm}{\centerline{
  \begin{fmfgraph*}(6,6)
  \setval
  \fmfforce{0.w,0.5h}{v1}
  \fmfforce{1w,0.5h}{v2}
  \fmf{plain}{v1,v2}
  \fmfv{decor.size=0, label=${\scs 1}$, l.dist=1mm, l.angle=-180}{v1}
  \fmfv{decor.size=0, label=${\scs 2}$, l.dist=1mm, l.angle=0}{v2}
  \end{fmfgraph*} } }
\hspace*{5mm}
\equiv \hspace*{2mm} G(\tau_1, \tau_2)
\label{PROP1}
\end{eqnarray}
defined by the identity
\begin{eqnarray}
\label{PROP2}
\int_0^\beta d \tau_2 \,G^{-1}(\tau_1 , \tau_2 ) \,G (\tau_2 , \tau_3 ) = \frac{1}{D} \,\delta  (\tau_1 - \tau_3 ) 
\end{eqnarray}
and the vertices
\begin{eqnarray}
%
  \parbox{7mm}{\centerline{
  \begin{fmfgraph*}(6,6)
  \setval
  \fmfforce{0w,0.5h}{v1}
  \fmfforce{0.5w,1h}{v2}
  \fmfforce{1w,0.5h}{v3}
  \fmfforce{0.5w,0h}{v4}
  \fmfforce{1/2w,1/2h}{v5}
  \fmf{plain}{v1,v3}
  \fmf{plain}{v2,v4}
  \fmf{dots,left=0.4}{v3,v4}
  \fmfv{decor.size=0, label=${\scs 1}$, l.dist=1mm, l.angle=-180}{v1}
  \fmfv{decor.size=0, label=${\scs 2}$, l.dist=1mm, l.angle=90}{v2}
  \fmfv{decor.size=0, label=${\scs 3}$, l.dist=1mm, l.angle=0}{v3}
  \fmfv{decor.size=0, label=${\scs n}$, l.dist=1mm, l.angle=-90}{v4}
  \fmfdot{v5}
  \end{fmfgraph*} } }
\hspace*{5mm}
\equiv -  D \,\int_0^{\beta}d\tau_1 \int_0^{\beta}d\tau_2  \int_0^{\beta}d\tau_3
\,\ldots \, \int_0^{\beta}d\tau_n
\left. \frac{ \delta^n{\cal A}[\sigma] }{\delta \sigma(\tau_1)
  \delta \sigma(\tau_2) \delta \sigma(\tau_3) \ldots  \delta \sigma(\tau_{n})} \right|_{\sigma(\tau)=\sigma_0}
\label{FeynRule3BMPI} \,.
\end{eqnarray}
For instance, the two-loop contribution is given by the Feynman diagrams
\begin{eqnarray}
V^{(2)}_{\rm eff} (\sigma_0)= \frac{1}{8}
\parbox{11mm}{\centerline{
\begin{fmfgraph}(8,4)
\setval
\fmfleft{i1}
\fmfright{o1}
\fmf{plain,left=1}{i1,v1,i1}
\fmf{plain,left=1}{o1,v1,o1}
\fmfdot{v1}
\end{fmfgraph}}}
+\frac{1}{12}
\parbox{7mm}{\centerline{
\begin{fmfgraph}(4,4)
\setval
\fmfforce{0w,0.5h}{v1}
\fmfforce{1w,0.5h}{v2}
\fmf{plain,left=1}{v1,v2,v1}
\fmf{plain}{v1,v2}
\fmfdot{v1,v2}
\end{fmfgraph}}}\,.
\label{EFF2}
\end{eqnarray}
In order to evaluate (\ref{EFF2}), we need the third and fourth functional derivatives
of the Euclidean action (\ref{AC2}). They are given by
\begin{eqnarray}
\left. \frac{ \delta^3{\cal A}[\sigma] }{\delta \sigma(\tau_1)
\delta \sigma(\tau_2) \delta \sigma(\tau_3)} \right|_{\sigma(\tau)=\sigma_0}
= - 64 i \,
G_\Omega ( \tau_1,\tau_2) G_\Omega ( \tau_2,\tau_3) G_\Omega ( \tau_3,\tau_1)
\end{eqnarray}
and
\begin{eqnarray}
&& \left. \frac{ \delta^4{\cal A}[\sigma] }{\delta \sigma(\tau_1)
\delta \sigma(\tau_2) \delta \sigma(\tau_3) \delta \sigma(\tau_4)} \right|_{\sigma(\tau)=\sigma_0} = - 256
\Big[ G_\Omega ( \tau_1,\tau_2) G_\Omega ( \tau_2,\tau_3) G_\Omega ( \tau_3,\tau_4) G_\Omega ( \tau_4,\tau_1) \nonumber \\
&& + G_\Omega ( \tau_1,\tau_2) G_\Omega ( \tau_2,\tau_4) G_\Omega ( \tau_4,\tau_3) G_\Omega ( \tau_3,\tau_1)
+ G_\Omega ( \tau_1,\tau_3) G_\Omega ( \tau_3,\tau_2) G_\Omega ( \tau_2,\tau_4) G_\Omega ( \tau_4,\tau_1)  \Big]  \, ,
\end{eqnarray}
where the explicit form of the correlation function $G_\Omega (\tau_1,\tau_2 )$ 
at zero temperature follows from (\ref{CF})  and (\ref{CON}):
\begin{eqnarray}
G_\Omega (\tau_1,\tau_2 ) = \frac{1}{2 \Omega}\,e^{-\Omega| \tau_1 - \tau_2 |} \, .
\end{eqnarray}
Furthermore, we have to solve (\ref{PROP2}) for the propagator (\ref{PROP1}).
Performing at arbitrary temperature the decomposition
\begin{eqnarray}
\label{GCOM}
G (\tau_1,\tau_2) = \frac{1}{\beta}\sum_{m=-\infty}^\infty G_m \, e^{-i \omega_m (\tau_1 - \tau_2 )} \, ,
\end{eqnarray}
the coefficient $G_m$ turns out to be
\begin{eqnarray}
\label{GG}
G_m = \frac{1}{D \,G_m^{-1}} \, .
\end{eqnarray}
Using (\ref{MCC}) and (\ref{GG}), we evaluate the Matsubara sum (\ref{GCOM})
at zero temperature according to (\ref{CON}) and obtain
\begin{eqnarray}
\label{GRES}
G(\tau_1,\tau_2) = \frac{\tilde{g}}{2 D} \left[ \delta (\tau_1 - \tau_2 ) - \frac{2 \tilde{g}}{\Omega \tilde{\Omega}}
\, e^{- \tilde{\Omega}| \tau_1 - \tau_2 |} \right] \, .
\end{eqnarray}
From (\ref{FeynRule3BMPI}) we read off that each vertex is of order $D$, whereas each
propagator is of order $1/D$ due to
(\ref{GRES}). Thus both Feynman diagrams in (\ref{EFF2}) are, indeed, of order $1/D$.
The first and second Feynman diagram in (\ref{EFF2}) lead to the expressions
\begin{eqnarray}
\label{V21}
V_{\rm eff}^{(2,1)} (\sigma_0) &=& \frac{1}{\beta \,D} \left\{
- \frac{\tilde{g}^2}{2 \Omega^4} \Big[ 2 I_2(2 \Omega) + I_2(4 \Omega) \Big]
+ \frac{2 \tilde{g}^3}{\Omega^5 \tilde{\Omega}}
\Big[ 2 I_3 ( \Omega, \Omega,\Omega+\tilde{\Omega}) + I_3 ( 2 \Omega, 2 \Omega,\tilde{\Omega}) \Big] \right. \nonumber \\
&& \left. - \frac{2 \tilde{g}^4}{\Omega^6 \tilde{\Omega}^2}
\Big[ 2 I_4(\Omega+\tilde{\Omega}, 0 , \Omega,\Omega,0,\Omega+\tilde{\Omega})
+ I_4(\tilde{\Omega}, \Omega , \Omega,\Omega,\Omega,\tilde{\Omega}) \Big] \right\} \, , \\
V_{\rm eff}^{(2,2)} (\sigma_0) &=& \frac{1}{\beta \,D} \left[
\frac{2 \tilde{g}^3}{3 \Omega^6} \, I_3(2 \Omega, 2 \Omega, 2 \Omega)
- \frac{4 \tilde{g}^4}{\Omega^7 \tilde{\Omega}}\, I_4 (\Omega, \Omega, \tilde{\Omega}, 2 \Omega, \Omega,\Omega)
+ \frac{8 \tilde{g}^5}{\Omega^8 \tilde{\Omega}^2}\,I_5(\Omega,\Omega,\tilde{\Omega},0,\Omega,0,\tilde{\Omega},
\Omega,\Omega,\Omega) \right. \nonumber \\
&&\left. - \frac{16 \tilde{g}^6}{3 \Omega^9 \tilde{\Omega^3}}\,I_6(\Omega,\Omega,\tilde{\Omega},0,0,\Omega,0,\tilde{\Omega},
0,0,0,\tilde{\Omega},\Omega,\Omega,\Omega) \right] \,  ,
\label{V22}
\end{eqnarray}
respectively. Here we have introduced an abbreviation for multiple integrals with
respect to imaginary times:
\begin{eqnarray}
I_n (\Omega_{12},\ldots,\Omega_{1n},\Omega_{23},\ldots,\Omega_{2n},\ldots) = \int_0^\beta d \tau_1 \int_0^\beta d \tau_2 \,\ldots \, \int_0^\beta d \tau_n \,
\exp \left( - \sum_{i=1}^n \sum_{j=i+1}^n \Omega_{ij}| \tau_i - \tau_j | \right) \, .
\end{eqnarray}
Considering the low-temperature limit $\beta \to \infty$, the first three of these integrals
read in closed form \cite{Diplomarbeit}:
\begin{eqnarray}
&& \hspace*{-1cm} I_2 (\Omega_{12}) = \frac{2\beta}{\Omega_{12}} \, , \\
&& \hspace*{-1cm} I_3 (\Omega_{12},\Omega_{13},\Omega_{23}) = \frac{4\beta(\Omega_{12}+\Omega_{13}+\Omega_{23})}{(\Omega_{12}
+\Omega_{13})(\Omega_{12}+\Omega_{23})(\Omega_{13}+\Omega_{23})} \, , \\
&& \hspace*{-1cm} I_4 (\Omega_{12},\Omega_{13},\Omega_{14},\Omega_{23},\Omega_{24},\Omega_{34}) = 2 \beta \nonumber \\
&& \hspace*{-1cm}\times \left\{ \frac{1}{\Omega_{12}+\Omega_{13}+ \Omega_{24}+ \Omega_{34}}
\left[ \frac{1}{\Omega_{12}+ \Omega_{13}+\Omega_{14}}+
\frac{1}{\Omega_{14}+ \Omega_{24}+\Omega_{34}} \right]
\left[ \frac{1}{\Omega_{12}+ \Omega_{23}+\Omega_{24}}
+ \frac{1}{\Omega_{13}+ \Omega_{23}+\Omega_{34}} \right] \right. \nonumber \\
&& \hspace*{-1cm}
+ \frac{1}{\Omega_{12}+\Omega_{14}+ \Omega_{23}+ \Omega_{34}}
\left[ \frac{1}{\Omega_{12}+ \Omega_{23}+\Omega_{24}}+
\frac{1}{\Omega_{14}+ \Omega_{24}+\Omega_{34}} \right]
\left[ \frac{1}{\Omega_{12}+ \Omega_{13}+\Omega_{14}}+
\frac{1}{\Omega_{13}+ \Omega_{23}+\Omega_{34}} \right]\nonumber \\
&& \hspace*{-1cm}\left.
+ \frac{1}{\Omega_{13}+\Omega_{14}+ \Omega_{23}+ \Omega_{24}}
\left[ \frac{1}{\Omega_{13}+ \Omega_{23}+\Omega_{34}}+
\frac{1}{\Omega_{14}+ \Omega_{24}+\Omega_{34}} \right]
\left[ \frac{1}{\Omega_{12}+ \Omega_{13}+\Omega_{14}}+
\frac{1}{\Omega_{12}+ \Omega_{23}+\Omega_{24}} \right] \right\} \, .
\end{eqnarray}
Furthermore, in order to evaluate ({\ref{V22}}), we need one particular integral
with respect to five and six imaginary times, respectively:
\begin{eqnarray}
\label{I5}
&& I_5(\Omega,\Omega,\tilde{\Omega},0,\Omega,0,\tilde{\Omega},
\Omega,\Omega,\Omega) = \frac{\beta (3 \tilde{\Omega}^5+ 42 \tilde{\Omega}^4\Omega + 227 \tilde{\Omega}^3 \Omega^2
+ 568 \tilde{\Omega}^2 \Omega^3 + 656 \tilde{\Omega} \Omega^4+ 288 \Omega^5)}{2 \Omega^2 (\tilde{\Omega}+ \Omega)^2
(\tilde{\Omega}+ 2 \Omega)^4 (\tilde{\Omega}+ 4 \Omega) } \, , \\
&&I_6(\Omega,\Omega,\tilde{\Omega},0,0,\Omega,0,\tilde{\Omega},
0,0,0,\tilde{\Omega},\Omega,\Omega,\Omega)  = \frac{3 \beta (\tilde{\Omega}^5+ 14 \tilde{\Omega}^4\Omega + 73 \tilde{\Omega}^3 \Omega^2
+ 160 \tilde{\Omega}^2 \Omega^3 + 136 \tilde{\Omega} \Omega^4+ 32 \Omega^5)}{ \tilde{\Omega} \Omega^2 (\tilde{\Omega}+ \Omega)^2
(\tilde{\Omega}+ 2 \Omega)^4 (\tilde{\Omega}+ 4 \Omega) } \, .
\label{I6}
\end{eqnarray}
From (\ref{L0}) and (\ref{V21})--(\ref{I6}) we read off the effective
potential for zero temperature up to the order $1/D$:
\begin{eqnarray}
\hspace*{-0.4cm}
V_{\rm eff}(\sigma_0) &=& D \left( \frac{\sigma_0^2}{\tilde{g}} + \frac{\Omega}{2} \right)
+ \frac{\tilde{\Omega}}{2}- \Omega
+ \frac{1}{D} \left[ - \frac{5 \tilde{g}^2}{4 \Omega^5}
+ \frac{\tilde{g}^3(\tilde{\Omega}^3 + 4 \tilde{\Omega}^2\Omega + 44 \tilde{\Omega}\Omega^2 + 128 \Omega^3)}{4 \Omega^8
\tilde{\Omega}(\tilde{\Omega}+2\Omega)^2}
\right.
\nonumber \\
&& - \frac{\tilde{g}^4 ( 3 \tilde{\Omega}^5+ 31 \tilde{\Omega}^4\Omega + 150 \tilde{\Omega}^3 \Omega^2
+ 392 \tilde{\Omega}^2 \Omega^3 + 656 \tilde{\Omega} \Omega^4+ 480 \Omega^5)}{ \Omega^{9} \tilde{\Omega}^2 (\tilde{\Omega}+ \Omega)
(\tilde{\Omega}+ 2 \Omega)^3 (\tilde{\Omega}+ 4 \Omega) }
\nonumber \\
&& + \frac{4 \tilde{g}^5 ( 3 \tilde{\Omega}^5+ 42 \tilde{\Omega}^4\Omega + 227 \tilde{\Omega}^3 \Omega^2
+ 568 \tilde{\Omega}^2 \Omega^3 + 656 \tilde{\Omega} \Omega^4+ 288 \Omega^5)}{ \Omega^{10} \tilde{\Omega}^2 (\tilde{\Omega}+ \Omega)^2
(\tilde{\Omega}+ 2 \Omega)^4 (\tilde{\Omega}+ 4 \Omega) } \nonumber \\
&& \left. - \frac{16 \tilde{g}^6(\tilde{\Omega}^5+ 14 \tilde{\Omega}^4\Omega + 73 \tilde{\Omega}^3 \Omega^2
+ 160 \tilde{\Omega}^2 \Omega^3 + 136 \tilde{\Omega} \Omega^4+ 32 \Omega^5)}{ \Omega^{11} \tilde{\Omega}^4 (\tilde{\Omega}+ \Omega)^2
(\tilde{\Omega}+ 2 \Omega)^4 (\tilde{\Omega}+ 4 \Omega) } \right]
+ {\cal O} \left( \frac{1}{D^2} \right) \, .
\label{VEFF}
\end{eqnarray}
The ground-state energy of the anharmonic oscillator (\ref{V}) is found by extremizing the
effective potential (\ref{VEFF}) with respect to the background $\sigma_0$ while taking into
account the auxiliary frequencies (\ref{FR}) and (\ref{AUX2}).
\subsection{Weak-Coupling}
In order to cross-check our large-$D$ result, we specialize now to
the weak-coupling regime where the extremal background is expanded according to
\begin{eqnarray}
\label{WE1}
\sigma_0 = - i \left( s_1 \tilde{g} + s_2 \tilde{g}^2 + s_3 \tilde{g}^3 + s_4 \tilde{g}^4
+ s_5 \tilde{g}^5 + \ldots \right) \, .
\end{eqnarray}
Inserting (\ref{WE1}) into the vanishing first derivative of (\ref{VEFF}) and re-expanding
in $\tilde{g}$, we obtain a system of equations which are solved by
\begin{eqnarray}
\label{WE2}
s_1 & = & \frac{1}{2 \omega}  \, , \\
s_2 & = & -\frac{1}{2 \omega^4} - \frac{1}{\omega^4 D} \, , \\
s_3 & = &\frac{5}{4 \omega^7} + \frac{45}{8 \omega^7 D} + \frac{25}{4 \omega^7 D^2}  \, , \\
s_4 & = & - \frac{4}{\omega^{10}} - \frac{59}{2 \omega^{10} D} - \frac{73}{\omega^{10} D^2} \, , \\
s_5 & = & \frac{231}{16 \omega^{13}} + \frac{19503}{128 \omega^{13} D} + \frac{9823}{16 \omega^{13} D^2}
+ \frac{275}{2 \omega^{13} D^3}  \, ,
\hspace*{0.4cm} \ldots \, .
\label{WE3}
\end{eqnarray}
Inserting (\ref{WE1})--(\ref{WE3}) into (\ref{VEFF}) and re-expanding in
$\tilde{g} = Dg$ yields again the weak-coupling series (\ref{AnsE0D}),
where the weak-coupling coefficients $\epsilon_k$ of Table \ref{EpsTableD} are reproduced in the first
three terms of order $D^{k+1}$, $D^k$, and $D^{k-1}$.
\subsection{Strong-Coupling}
\begin{figure}[t]
\centerline{\includegraphics[scale=0.65]{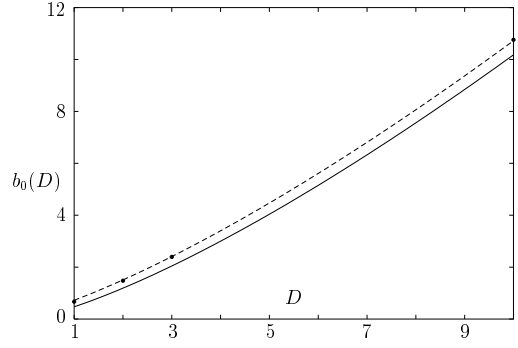}}
\caption{Strong-coupling coefficient $b_0(D)$ versus dimension $D$. Solid
and dashed lines represent leading and subleading results of the large-$D$
expansion (\ref{ST3})--(\ref{ST4B}), respectively.
For $D=1$, the dot represents the earlier result from Refs.~\cite{JankeC1,JankeC2};
for $D=2,\,3,\,10$, the dots indicate the highly accurate VPT values
obtained from the 80th order, as given in Table \ref{ResultsD}.}
\label{F2}
\end{figure}

We now derive the large-$D$ expansion for the ground-state energy in the
strong-coupling regime. There,
we use for the extremal background the ansatz
\begin{eqnarray}
\label{ST1}
\sigma_0 = - i \tilde{g}^{2/3} \left( S_1  + S_2 \tilde{g}^{-2/3} + \ldots \right) 
\end{eqnarray}
and find for the leading coefficient the expansion
\begin{eqnarray}
\label{ST2}
S_1 = \frac{1}{2^{4/3}} + \left( 2^{1/6}\cdot 3^{1/2} - 2^{5/3} \right)
\frac{1}{6 D}  + \left( 27 \cdot 6^{1/2}- \frac{73}{2} \right)^{1/3} \frac{1}{18 D^2}
+ {\cal O} \left( \frac{1}{D^3} \right) \, .
\end{eqnarray}
From (\ref{VEFF}), (\ref{ST1}), and (\ref{ST2}) we then obtain the strong-coupling
series (\ref{EX2}) where
the leading strong-coupling coefficient has the large-$D$ expansion
\begin{eqnarray}
\label{ST3}
b_0 = \sum_{k=0}^\infty B_k\,D^{4/3-k},
\end{eqnarray}
with
\begin{eqnarray}
\label{ST4A}
B_0 &=& \frac{3 \cdot 2^{1/3}}{8} \, , \\
\label{ST4B}
B_1 &=& \frac{3^{1/2}-2^{1/2}}{2^{1/6}} \approx 0.2831607943221791188446646047948820365123\, , \\
\label{ST4C}
B_2 & = & - \frac{239}{18 \cdot 2^{2/3} \left( 25 + 12 \cdot6^{1/2} \right)}
\approx  -0.1537760559399284913195761085499705701590 \, .
\end{eqnarray}
Figure \ref{F2} shows the leading strong-coupling coefficient (\ref{ST3})--(\ref{ST4B})
up to the first two orders plotted as a function of the dimension $D$.\\
\section{Large-$D$ Expansion From VPT}
In this section, we study how the large-$D$ expansion (\ref{ST3}) follows from VPT. Thereby,
we numerically determine the large-$D$ expansion coefficients up to $B_6$ with high precision.
\subsection{Coefficients $B_1$ and $B_2$}
\begin{figure}[t]
\includegraphics[scale=0.65]{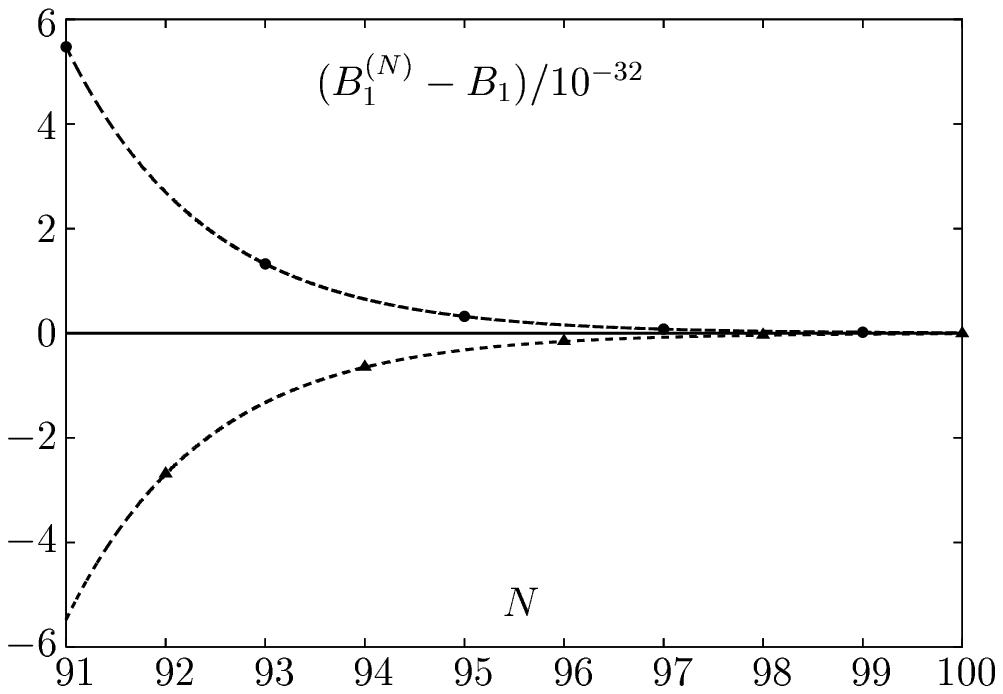}
\hspace{15mm}
\includegraphics[scale=0.65]{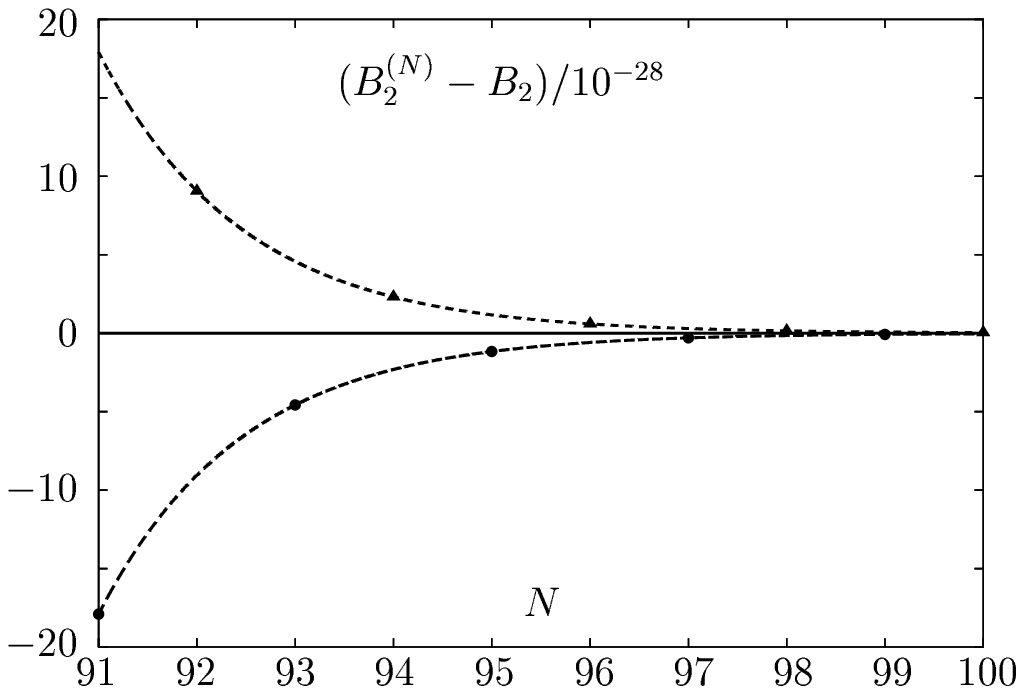}
\caption{Deviation of the VPT approximants for $B_1^{(N)}$ and $B_2^{(N)}$ from their
exact values $B_1$ and $B_2$ in Eqs.~(\ref{ST4B}), (\ref{ST4C}), respectively. Odd
orders are represented by circles; even orders by triangles.  The dashed lines
represent fits of the data to exponential functions.}
\label{F34}
\end{figure}

To first order, VPT gives (\ref{b1D}),
and we obtain the expansion (\ref{ST3}) simply by expanding the
cubic root in (\ref{b1D}) in powers of $1/D$.  The first three
coefficients in the expansion (\ref{ST3}) read
\begin{eqnarray}
\label{ST5}
B_0^{(1)} =  \frac{3 \cdot 2^{1/3}}{8}  \, , \hspace*{0.5cm} B_1^{(1)} = \frac{2^{1/3}}{4}
\, , \hspace*{0.5cm} B_2^{(1)} = -\frac{2^{1/3}}{6} \,.
\end{eqnarray}
The leading coefficient (\ref{ST4A}) is reproduced exactly,
while the next two subleading coefficients $B_1$ and $B_2$ are missed by 11.2 \% and 36.6 \%,
respectively. In the second order of VPT, Eq.~(\ref{b0Nred}) yields
\begin{eqnarray}
b_0^{(2)}(\Omega_0^{(2)}) = \frac{3D\Omega_0^{(2)}}{16} + \frac{D(2 + D)}{2(\Omega_0^{(2)})^2} -
\frac{D(10 + 9D + 2D^2)}{ 8 (\Omega_0^{(2)})^5 } \, , 
\label{b02}
\end{eqnarray}
which has to be extremized with respect to the variational parameter $\Omega_0^{(2)}$.
Setting the derivative of (\ref{b02}) to zero, the large-$D$ expansion
\begin{eqnarray}
\label{varexp2}
\Omega_0^{(2)} = D^{1/3}\left(C_0^{(0,2)} +
\frac{C_1^{(0,2)}}{D} + \frac{C_2^{(0,2)}}{D^2} + \ldots \right)
\end{eqnarray}
leads to a system of equations for the expansion coefficients whose solutions read
\begin{eqnarray}
\label{optcoe}
C_0^{(2)} = 2^{1/3} \, , \hspace*{0.5cm}
C_1^{(2)} = \frac{13 \cdot 2^{1/3}}{12} \, , \hspace*{0.5cm}
C_2^{(2)} = -\frac{113 \cdot 2^{1/3}}{288} \, , \hspace*{0.5cm} \ldots \, .
\end{eqnarray}
Thus, by inserting the optimized variational parameter (\ref{varexp2}), (\ref{optcoe})
into (\ref{b02}) and expanding in $1/D$ we obtain the first three coefficients
in (\ref{ST3}) to second order of VPT:
\begin{eqnarray}
\label{ST6}
B_0^{(2)} = \frac{3 \cdot 2^{1/3}}{8} \, , \hspace*{0.5cm} B_1^{(2)} =
\frac{7 \cdot 2^{1/3}}{32} \, , \hspace*{0.5cm} B_2^{(2)} = - \frac{71 \cdot 2^{1/3}}{768} \, .
\end{eqnarray}
The leading large-$D$ coefficient $B_0$ remains the same, whereas the error of
the subleading coefficient $B_1^{(2)}$ is reduced to 2.67 \% and the next subleading
coefficient $B_2^{(2)}$ now comes out with an error of 24.2 \%.
Figure \ref{F34} shows that the VPT approximants $B_1^{(N)}$ and $B_2^{(N)}$  converge
exponentially fast to their exact values $B_1$ and $B_2$ in (\ref{ST4B}) and (\ref{ST4C}).
If we had not known the exact analytic result for the subleading coefficients $B_1$ and $B_2$,
we could have extracted its value from the VPT approximants as follows.  Figure \ref{F34}
shows that the VPT approximants for odd and even orders converge towards the exact value
independently. Extrapolating the values of $B_1^{(\rm odd)}$ and $B_1^{(\rm even)}$ for
$N \to \infty$ leads to the limiting values
\begin{eqnarray}
B_1^{({\rm odd})}  \approx 0.28316079432217911884466460479488203808\, , \\
B_1^{({\rm even})} \approx 0.28316079432217911884466460479488203575\, .
\end{eqnarray}
Assuming that the exact value lies within this interval,
we obtain from this extrapolation method the result
\begin{eqnarray}
B_1^{({\rm extrap})} = 0.2831607943221791188446646047948820369(24) \, .
\end{eqnarray}
An analogous procedure can be applied to extract a numerical value for
the subsequent coefficient $B_2$.
Applying the extrapolation method for odd and even orders, we obtain the result
\begin{eqnarray}
B_2^{({\rm extrap})} = -0.153776055939928491319576108549961(60) \, .
\end{eqnarray}
\subsection{Coefficients up to $B_6$}
\begin{figure}[t]
\centerline{\rotatebox{-0}{\includegraphics[scale=0.65]{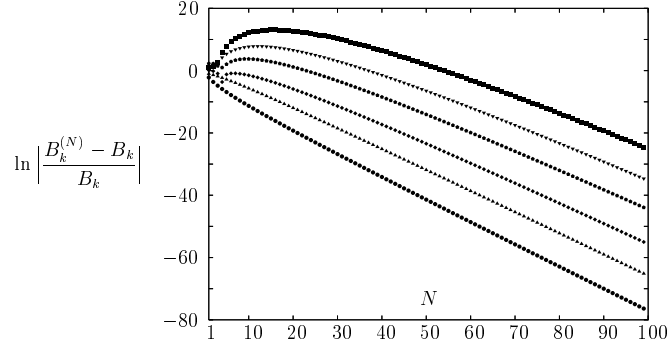}}}
\caption{Logarithm of the relative deviation of the VPT results $B_k^{(N)}$
for the coefficients $B_1$ through $B_6$ in the large-$D$ expansion (\ref{ST3})
versus the order $N$ of VPT. The lowest curve (circles) is for $B_1^{(N)}$; the
uppermost curve (squares) is for $B_6^{(N)}$; intermediate curves are for
$B_2^{(N)}$ through $B_5^{(N)}$ ($B_2^{(N)}$: triangles, $B_3^{(N)}$: diamonds,
$B_4^{(N)}$: pentagons, $B_5^{(N)}$: upside-down triangles).  For $B_1^{(N)}$ and
$B_2^{(N)}$ the exact values from (\ref{ST4B}) and (\ref{ST4C}) were used in order
to determine the relative deviation.  For $B_3^{(N)}$ through $B_6^{(N)}$ we used
our extrapolation results from Table \ref{T4}.}
\label{F5}
\end{figure}

Figure \ref{F5} shows that the VPT approximants for $B_1$ and $B_2$ rapidly converge
towards their exact values.  For the large-$D$ expansion coefficients of higher orders
in $1/D$, however, the VPT approximants first fluctuate and then enter the regime of
exponentially fast convergence.  If we want to obtain the coefficients of the
expansion (\ref{ST3}) to higher orders and with good accuracy, we must therefore drive
our VPT calculation to high orders.  To this end, we specialize the expression
for the leading strong-coupling coefficient as given in (\ref{b0Nred}) in
such a way that we can read off the corresponding large-$D$ expansion.
The weak-coupling coefficients for the ground-state energy $\epsilon_k$
can be expanded in powers of the spatial dimension $D$,
\begin{eqnarray}
\epsilon_k = \sum_{j = 1}^{k + 1} \epsilon_j^{(k)}D^j \, , \label{epskD}
\end{eqnarray}
whereas the variational parameter $\Omega_0^{(N)}$ can be expanded in $1/D$:
\begin{eqnarray}
\Omega_0^{(N)} = D^{1/3}\sum_{j = 0}^{M}C_j^{(0,N)}D^{-j}\, . \label{Om0D}
\end{eqnarray}
Here, $M$ denotes the highest order to which we seek to drive the
large-$D$ expansion (\ref{ST3}).
Furthermore, recall the multinomial expansion
\begin{eqnarray}
(x_1 + x_2 + \ldots + x_m)^n = \sum_{a_1 = 0}^n \sum_{a_2 = 0}^n \ldots
\sum_{a_m = 0}^n \delta\left(n, \sum_{k = 1}^m a_k \right) \frac{n!}{a_1!\,  a_2!\,  \cdots a_m!}
\prod_{l = 1}^m x_l^{a_l}\, ,
\end{eqnarray}
where $n$ is an integer and where the Kronecker delta $\delta_{ij}$ is written as $\delta(i,j)$.
This multinomial expansion can be generalized to hold for real exponents and infinite series
as follows:
\begin{eqnarray}
(1 + x_1 + x_2 + \ldots )^{\alpha} &=& \sum_{m = 0}^{\infty} \left[ \sum_{a_1 = 0}^{\infty} \sum_{a_2 = 0}^{\infty} \ldots
\sum_{a_m = 0}^{\infty} \delta\left(m,\sum_{l = 1}^m l a_l \right) \right. \nonumber \\
&& \left. \times \,\frac{\Gamma(\alpha + 1)  }{\Gamma(\alpha - a_1 - a_2 - \ldots - a_m + 1) a_1!
\,  a_2!\,  \ldots a_m!} \, {\displaystyle \prod_{l = 1}^{m}x_l^{a_l}}\right]  \, .
\label{MON}
\end{eqnarray}
Using (\ref{epskD}) and (\ref{Om0D}) in (\ref{b0Nred}) and
applying the multinomial expansion (\ref{MON}), we obtain
\begin{eqnarray}
b_0^{(N)}(\Omega_0^{(N)}) &=& D^{1/3} \left\{ \frac{D}{2}\sum_{j = 0}^M C_j^{(0,N)}D^{-j} + \, \,
\sum_{k =1}^N {(1 -3 k )/2 - 1 \choose N - k}
(-1)^{N-k}D^{- k}\left(C_0^{(0,N)}\right)^{1 - 3k} \right. \nonumber \\
&& \times \left. \left[  \sum_{j=1}^{k+1} \epsilon_j^{(k)} D^j+
\sum_{j=2}^{M + k} \,\, \sum_{m = 1}^{{\rm Min} \left\{ j - 1, \, \lfloor (M + j - k)/2\rfloor
- 1 \right \}} \epsilon_{j - m}^{(k)}  K_{m,k}^{(N)}  D^{j-2m} \right] \right\} \,,
\label{b0Nopt}
\end{eqnarray}
where the coefficients $K_{m,k}^{(N)}$ are given by
\begin{eqnarray}
K_{m,k}^{(N)} = \sum_{a_1 = 0}^M \sum_{a_2 = 0}^M \ldots \sum_{a_m = 0}^M
\delta\left(m, \sum_{l = 1}^{m}la_l\right)\frac{(3k - 2 + a_1 + a_2 + \ldots a_m)!}{(3k - 2)!\,
a_1!\,  a_2!\,  \ldots a_m!}(-1)^{a_1 + a_2 + \ldots + a_m}\prod_{l = 1}^{m}
\left(\frac{C_l^{(0,N)}}{C_0^{(0,N)}}\right)^{a_l}\,. \label{Kmk}
\end{eqnarray}
The summation boundaries in (\ref{b0Nopt}) are determined by the two conditions
that i) $\epsilon_j^{(k)} = 0$ for $j <1 \lor j > k+1$ and ii) we can neglect contributions
containing coefficients $C_k^{(0,N)}$ with $k > M$. Note that a similar expansion holds
for the derivative of the leading strong-coupling coefficient with respect to the
variational parameter $d \, b_0^{(N)} / d \, \Omega_0^{(N)}$.
Using  (\ref{b0Nopt}), (\ref{Kmk}) we can efficiently calculate the
VPT approximants to high orders.  In each order of VPT, the leading large-$D$ coefficient
$B_0$ comes out exactly. Furthermore, the expansion coefficients $C_k^{(0,N)}$ for the
variational parameter are found by solving linear equations. Figure \ref{F5} shows
the convergence behavior of our VPT approximants for $B_1$ through $B_6$ up to $N=100$.
By extrapolating the VPT results $B_k^{(N)}$ for $k \geq 3$ in the limit $N \to \infty$,
we are able to determine the coefficients in the large-$D$ expansion to high
accuracy. The numerical results are shown in Table \ref{T4}.
\section{Summary}
We have determined the weak-coupling series of the
ground-state energy for the $D$-dimensional anharmonic oscillator (\ref{V})
and used variational perturbation theory
to extract the coefficients of the
large-$D$ expansion to higher orders than accessible
by standard field-theoretic methods
based on the Hubbard-Stratonovich transformation.
\section*{Acknowledgement}
We cordially thank Hagen Kleinert for having inspired the present work and for carefully reading
the manuscript.
\begin{table}[h]
\begin{center}
\begin{tabular}{|l|l|}
\hline
\rule[-1mm]{0mm}{4.5mm}\hspace*{1mm}$B_3$\hspace*{1mm}  &
\hspace*{1mm}0.1098507701648367991179224418(38)\hspace*{1mm}  \\[1mm] \hline
\rule[-1mm]{0mm}{4.5mm}\hspace*{1mm}$B_4$\hspace*{1mm}  &
\hspace*{1mm}-0.02886373198599697649759(11)\hspace*{1mm}  \\[1mm] \hline
\rule[-1mm]{0mm}{4.5mm}\hspace*{1mm}$B_5$\hspace*{1mm}  &
\hspace*{1mm}-0.1695976321261828993(92)\hspace*{1mm}  \\[1mm] \hline
\rule[-1mm]{0mm}{4.5mm}\hspace*{1mm}$B_6$\hspace*{1mm}  &
\hspace*{1mm}0.51902738902696(86)\hspace*{1mm}  \\[1mm] \hline
\end{tabular}
\caption{VPT results for the coefficients in the large-$D$ expansion (\ref{ST3}) of the leading strong-coupling coefficient $b_0$.}
\label{T4}
\end{center}
\end{table}
\end{fmffile}
\end{document}